\documentclass[useAMS,usenatbib,usegraphicx,fleqn]{mn2e}
\usepackage[english]{babel}
\usepackage[latin1]{inputenc}
\usepackage{amsmath}
\usepackage{amsfonts}
\usepackage{amssymb}
\usepackage[]{graphicx}
\usepackage{verbatim}
\usepackage{mathrsfs}
\usepackage{enumerate} 
\usepackage{paralist}
\usepackage{color}

\newif\ifAMStwofonts
\AMStwofontstrue

\title[The IGIMF in dSphs]{The IGIMF and other IMFs in dSphs: the case of Sagittarius}
\author[F. Vincenzo et al.]{F. Vincenzo$^{1,2}$\thanks{E-mail:
vincenzo@oats.inaf.it}, F. Matteucci$^{1,2,3}$, S. Recchi$^{4}$, F. Calura$^{5}$, A. McWilliam$^{6}$, 
\newauthor G.~A. Lanfranchi$^{7}$
\\
$^{1}$Dipartimento di Fisica, Sezione di Astronomia, Universit\`a di Trieste, via G.B. Tiepolo 11, 34100, Trieste, Italy\\
$^{2}$INAF, Osservatorio Astronomico di Trieste, via G.B. Tiepolo 11, 34100, Trieste, Italy\\
$^{3}$INFN, Sezione di Trieste, Via Valerio 2, 34100, Trieste, Italy\\
$^{4}$Department of Astrophysics, Vienna University, T\"urkenschanzstrasse 17, A-1180, Vienna, Austria\\
$^{5}$INAF, Osservatorio Astronomico di Bologna, via Ranzani 1, I-40127, Bologna, Italy\\
$^{6}$The Observatories of the Carnegie Institute of Washington, 813 Santa Barbara Street, Pasadena, CA 91101, USA\\
$^{7}$N\'ucleo de Astrof\'isica Te\'orica, Universidade Cruzeiro do Sul, Rua Galv\~ao Bueno 868, 01506-000, S\~ao Paulo, Brazil}

\begin{document}

\date{Accepted 2015 February 18. Received 2015 February 16; in original form 2014 October 20}

\pagerange{\pageref{firstpage}--\pageref{lastpage}} \pubyear{2015}

\maketitle

\label{firstpage}

\begin{abstract}

We have studied the effects of various initial mass functions (IMFs) on the chemical evolution of the Sagittarius dwarf galaxy (Sgr). 
In particular, we tested the effects of the integrated galactic initial mass function (IGIMF) on various predicted 
abundance patterns. 
The IGIMF depends on the star formation rate and metallicity and predicts less 
massive stars in a regime of low star formation, as it is the case in dwarf spheroidals.
We adopted a detailed chemical evolution model following the evolution of $\alpha$-elements, Fe and Eu, and assuming the currently best set of stellar yields. 
We also explored different yield prescriptions for the Eu, including production from neutron star mergers. 
Although the uncertainties still present in the stellar yields and data prevent us from drawing firm conclusions, our results suggest that the IGIMF applied to Sgr predicts lower [$\alpha$/Fe] 
ratios than classical IMFs and lower [hydrostatic/explosive] $\alpha$-element 
ratios, in qualitative agreement with observations.
In our model, the observed high [Eu/O] ratios in Sgr is due to reduced 
O production, resulting from the IGIMF mass cutoff of the massive  
oxygen-producing stars, as well as to the Eu yield produced in neutron star mergers, a more promising site than core-collapse supernovae, 
although many uncertainties are still present 
in the Eu nucleosynthesis. 
We find that a model, similar to our previous calculations, based on the late 
addition of iron from the Type~Ia supernova time-delay (necessary to reproduce the shape of [X/Fe] versus [Fe/H] relations) but 
also including the reduction of massive stars due to the IGIMF, better 
reproduces the observed abundance ratios in Sgr than models without the IGIMF.

\end{abstract}

\begin{keywords}
stars: abundances - galaxies: abundances - galaxies: dwarf - galaxies: evolution - galaxies: formation - Local Group. 
\end{keywords}

\section{Introduction} \label{intro}

The Sagittarius (Sgr) dwarf galaxy was the last classical dwarf spheroidal (dSph)  
discovered before the advent of the Sloan Digital Sky Survey. Its discovery was 
made by \citet{ibata1994} and it was identified while performing a spectroscopic radial 
velocity survey of the Galactic bulge stars \citep{ibata1997}. 
Its heliocentric distance ($D_{\sun}=26\pm2$ kpc, from \citealt{simon2011}) makes it 
the second closest known satellite galaxy of the Milky Way (MW) and, because of the strong tidal interaction
suffered by the Sgr dSph during its orbit, it has left behind a well-known stellar stream 
\citep{ibata2001,majewski2003,belokurov2006}, whose chemical characteristics have been
recently studied and compared with the ones of the Sgr main body and other 
dSph galaxies by \citet{deboer2014}. The Sgr dwarf galaxy has been classified 
as a dSph because of its very low central surface brightness 
($\mu_{\text{V}}=25.2\pm0.3$ mag arcsec$^{-2}$, from \citealt{majewski2003}), 
its very small total amount of gas ($M_{\text{HI,obs}}\sim10^4\,\text{M}_{\sun}$, from \citealt{mcconnachie2012}) 
and because of the age and metallicity of its main stellar population, 
which dates back to the age of the Universe and it is on average very iron poor.
Chemical abundances in Sgr have been measured by many authors up to now (see 
\citealt{lanfranchi2006} and references therein). Most of these studies have shown that 
the abundance patterns in Sgr are different than those in the MW.

\par  This work aims at testing the suggestions of \citet{mcwilliam2013},
which claimed that the $\alpha$-element deficiencies observed in the
Sgr dSph galaxy cannot be explained only by
means of the time-delay model \citep{tinsley1979,greggio1983,matteucci1986} but they rather
result from an initial mass function (IMF) deficient in the highest
mass stars. On the other hand, \citet{lanfranchi2003,lanfranchi2004} suggested that the low values 
of the [$\alpha$/Fe] ratios\footnote{We adopt
the following notation for the stellar chemical abundances:
$\text{[}X/Y\text{]}=\log_{10}(N_{X}/N_{Y})_{\star}-\log_{10}(N_{X}/N_{Y})_{\odot}$,
where $N_{X}$ and $N_{Y}$ are the volume density number of the atoms
of the species $X$ and $Y$, respectively, and we have scaled the stellar 
abundances to the solar abundances of \citet{asplund2009}.}, as observed
in dSphs, can be interpreted as due to the time-delay model ($\alpha$-elements
produced on short time-scales by core-collapse SNe and Fe by SNe Ia with a time delay) 
coupled with a low star formation rate (SFR), assuming a \citet{salpeter1955} IMF. 

\par \citet{mcwilliam2013} also suggested  that the Eu abundances in the Sgr galaxy 
might be explained by core-collapse SNe, whose progenitors are stars less massive than 
the main oxygen producers, as previously envisaged also by \citet{wanajo2003}. 
Since many studies of nucleosynthesis 
have pointed out the difficulty of producing r-process elements during SN explosions \citep{arcones2007}, 
in this work, we test different scenarios for the Eu production, both the 
one in which Eu is produced by core-collapse SNe and the most recent one, where the Eu is synthesized 
in neutron star mergers (NSMs, 
\citealt{korobkin2012,tsujimoto2014,shen2014,vandevoort2015}). 
The latter scenario was recently explored in the context of a detailed 
chemical evolution model by \citet{matteucci2014}, where they 
were able to well match the [Eu/Fe] abundance ratios observed in the MW stars. 

\par Here, we study the detailed chemical evolution
of Sgr by comparing the effect of the integrated galactic
initial mass function (IGIMF, in the formulation of
\citealt[hereafter R14]{recchi2014}) with the predictions of the canonical \citet{salpeter1955} 
and \citet{chabrier2003} IMFs. 
The main effect of the metallicity-dependent IGIMF of
R14 is a dependence of the maximum possible stellar mass, that can be formed
within a stellar cluster, on the [Fe/H] abundance and on the SFR, 
especially if the latter is very low. 
Since the dSphs turn out to have been characterized by very low SFRs, 
the effect of the IGIMF on their evolution is expected to be important. 
In this way, we will be able to test if the hypothesis of \citet{mcwilliam2013} 
is correct. 

\par In the past, the effect of the IGIMF in the chemical evolution of elliptical galaxies 
has been studied by \citet{recchi2009}, while \citet{calura2010} modelled the chemical evolution
of the solar neighbourhood when assuming the IGIMF. The originality of this work resides also
in the fact that we test the effect of a metallicity-dependent IMF, a study never done before in the 
framework of a detailed chemical evolution model. As a further element of originality, we include 
for the first time the Eu from NSMs in a chemical evolution model of a dSph galaxy.

\par Our work is organized as follows. In Section \ref{the_igimf}, we describe the metallicity-dependent 
IGIMF formalism that we include in our models. In Section \ref{data_sample}, the data sample is presented. 
In Section \ref{the_chemical_evolution_model}, we describe the chemical evolution model
adopted for the Sgr dSph and in Section \ref{results} we show and discuss
the results of our study. Finally, in Section \ref{conclusions},
we summarize the main conclusions of our work.

\section{The integrated galactic initial mass function} \label{the_igimf}   

Following \citet{kroupa2003} and \citet{weidner2005},the IGIMF is
defined by weighting the classical IMF, $\phi(m)$, with the mass
distribution function of the stellar clusters,
$\xi_{\text{ecl}}(M_{\text{ecl}})$, within which the star formation
process is assumed to take place: \begin{multline}
    \xi_{\text{IGIMF}}(m,\psi(t),\text{[Fe/H]})  \\
    = \int_{M_{\text{ecl,min}}}^{M_{\text{ecl,max}}(\psi(t))}{dM_{\text{ecl}}\,\xi_{\text{ecl}}(M_{\text{ecl}})\phi(m\leq
      m_{\text{max}},\text{[Fe/H]})}. \label{eq:igimf} \end{multline} 
The IGIMF is normalized in mass, such that
\[\int_{m_{\text{min}}}^{m_{\text{max}}}{dm\,m\,\xi_{\text{IGIMF}}(m,\psi(t),\text{[Fe/H]})}=1.\]
The functional form of the IGIMF that we test in this work depends
both on the SFR and on the [Fe/H] abundance of the parent galaxy,
following the \textit{mild model} of R14. The IGIMF is based on the following 
assumptions, based on observations. 

 \begin{enumerate}
 
 \item The mass spectrum of the embedded stellar clusters is assumed
   to be a power law, $\xi_{\text{ecl}}\propto
   M_{\text{ecl}}^{-\beta}$, with a slope $\beta=2$
   \citep{zhang1999,recchi2009}. In accordance with the mass of the
   smallest star-forming stellar cluster known (the Tauris---Auriga
   aggregate), we assumed $M_{\text{ecl, min}}=5\;\text{M}_{\sun}$, whereas
   the upper mass limit of the embedded cluster is a function of the
   SFR \citep{weidner2004}:
\begin{equation}
 \log{M_{\text{ecl,max}}} = A + B\,\log{\frac{\psi(t)}{\text{M}_{\sun}\text{yr}^{-1}}},
\label{eq:meclmax}
\end{equation} with $A=4.83$ and $B=0.75$.
 
\item Within each embedded stellar cluster of a given mass
  $M_{\text{ecl}}$ and [Fe/H] abundance, the IMF is assumed to be
  invariant. In our study, we assume an IMF which is defined as a
  two-slope power law:
\begin{equation}
\phi(m)= \left\{ \begin{array}{l l}
    A\,m^{-\alpha_{1}} & \quad \text{for $0.08\, \text{M}_{\sun} \leq m < 0.5\, \text{M}_{\sun}$}\\
    B\,m^{-\alpha_{2}} & \quad \text{for $0.5\, \text{M}_{\sun} \leq m < m_{\text{max}}$,}
  \end{array} \right. \label{eq:imf}
\end{equation} 
where $\alpha_{1}=1.30$ and $\alpha_{2}=2.35$ as in the original work of
\citet{weidner2005}, and $\alpha_{2}=2.3+0.0572\cdot\text{[Fe/H]}$ in the
mild formulation of R14. The latter relation was
  adapted by R14 from the original work of
  \citet{marks2012}.
It turns out from equation (\ref{eq:imf}) that the
overall [Fe/H] dependence entirely resides in the slope of the IMF
of the high-mass range. The maximum stellar mass $m_{\text{max}}$ that
can occur in the cluster and up to which the IMF is sampled, is
calculated according to the mass of the embedded cluster,
$M_{\text{ecl}}$; furthermore, $m_{\text{max}}$ must be in any case
smaller than the empirical limit, which here has been assumed to be 
$150\;\text{M}_{\sun}$ (see, for more details, \citealt{weidner2004}). The
$m_{\text{max}}\text{---}M_{\text{ecl}}$ relation is simply due to the fact
that, in the case of very low SFRs, the small clusters may not have
enough mass to give rise to very massive stars; on the other hand, in
the case of large SFRs, the maximum possible mass of the embedded
clusters may be very large and so very massive stars are able to
originate (see R14). So $m_{\text{max}}$ depends both on the
SFR and, to a lesser extent, on the [Fe/H] abundance of the parent 
galaxy.

\end{enumerate} It is worth remarking that recent studies
  \citep{weid2011,krou2013,weid2013} suggest that the star cluster IMFs can
  become top-heavy at SFRs larger than $\sim10$ M$_{\sun}$ yr$^{-1}$.
  This range of SFRs is clearly out of reach for Sgr,
  therefore we have neglected this modification of the IGIMF theory in
  our study.

\par
\begin{figure}
\includegraphics[width=9cm]{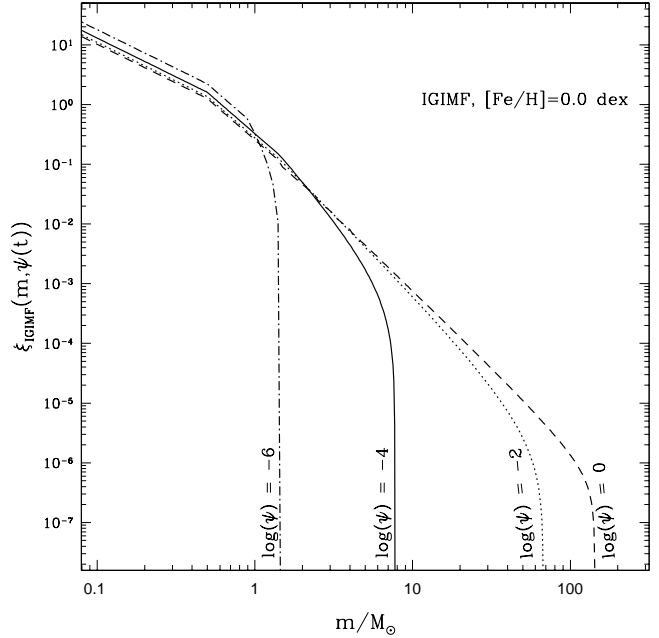}
\caption{In this figure, we show the predicted IGIMF, $\xi_{\text{IGIMF}}$,
  as a function of the stellar mass $m$ in the case of solar iron
  abundance and for various SFRs: from
  $\psi(t)=10^{-6}\,\text{M}_{\sun}$yr$^{-1}$ up to
  $\psi(t)=1\,\text{M}_{\sun}$yr$^{-1}$. The net effect of lowering the SFR
  is to truncate the IGIMF towards lower stellar masses.}
     \label{IGIMF}
   \end{figure}

 In Fig. \ref{IGIMF}, we show what is the effect of the dependence
of the IGIMF upon the SFR. For very low SFRs
($\leq1\,\text{M}_{\sun}$yr$^{-1}$), the IGIMF turns out to be very much
truncated and the maximum mass that can be formed strongly depends on
the SFR.  This is due to the fact that, in galaxies with a low SFR, 
the mass distribution function
of the embedded clusters is truncated at low values of
  $M_{\text{ecl}}$ (see equation \ref{eq:meclmax}) and small embedded
  clusters cannot produce very massive stars.

\par
\begin{figure}
\includegraphics[width=9cm]{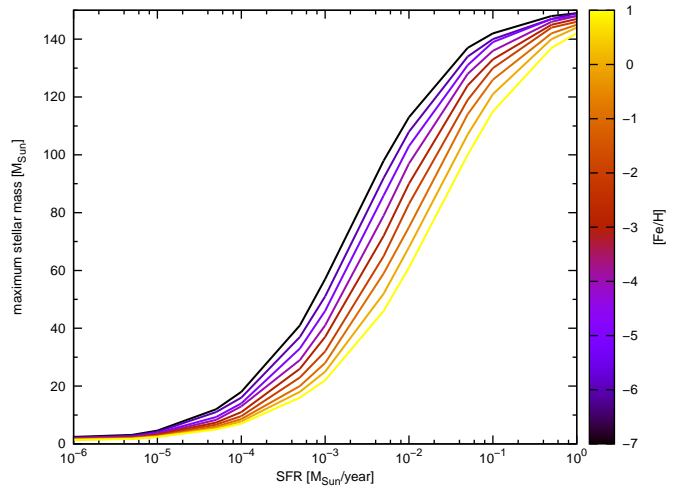}
\caption{This figure shows how the maximum stellar mass (on the
  $y$-axis) varies as a function of the SFR (on the $x$-axis) and as a
  function of the [Fe/H] abundance (colour-coding: from $-7$ dex up
  to $1$ dex). Increasing the [Fe/H] abundance has an opposite
  effect on $m_{\text{max}}$ with respect to the increasing of the
  SFR.}
     \label{m_max}
\end{figure}

 In Fig. \ref{m_max}, we illustrate how $m_{\text{max}}$ is
determined both by the SFR and by the [Fe/H] abundance. The effect of the
[Fe/H]-dependence is clearly opposite to that of the SFR-dependence.
As time passes by, one would expect that the iron content within the
galaxy insterstellar medium (ISM) increases; so, in this formulation,
if the SFR is constant, the maximum stellar mass that can be formed
within each embedded cluster is expected to decrease in time. 
In any case, it is worth noting that the
dependence of $m_{\text{max}}$ upon the SFR is much stronger than the
dependence upon [Fe/H] when the SFRs under play are extremely low.

\section{Data sample} \label{data_sample}

We use the data set of chemical abundances from the works of
\citet{bonifacio2000,bonifacio2004}, \citet{sbordone2007} 
and \citet{mcwilliam2013}.
\citet{bonifacio2000} derived the abundances of many
chemical elements for two giant stars in Sgr, observed with the 
high-resolution Ultraviolet and Visual Echelle Spectrograph (UVES) at the 
Kueyen-Very Large Telescope (VLT). \citet{bonifacio2004} did a similar work 
as \citet{bonifacio2000}, including the two stars previously analysed. From 
the former work we took the abundances of Eu, while from the latter one 
we took the abundances of Mg and O.
\citet{sbordone2007}
presented the chemical abundances of $12$ red giant stars belonging to
the Sgr main body and the chemical abundances of five red giant
stars belonging to the Sgr globular cluster Terzan 7, acquired
with the UVES at the 
European Southern Observatory (ESO) VLT.  
\citet{mcwilliam2013} derived the abundances of several
chemical elements from high-resolution spectra of three stars lying on
the faint red giant branch of M54, which is considered the most
populous globular cluster of Sgr, lying in the densest regions
of the galaxy. \citet{mcwilliam2013} acquired the spectra using the
Magellan Echelle spectrograph (MIKE) and their three stars were
confirmed from their kinematics to belong to the Sgr galaxy by
\citet{bellazzini2008}. \citet{mcwilliam2013} found their chemical 
abundances of Eu and Mg consistent with those of \citet{bonifacio2000,bonifacio2004}.
 
  \begin{figure}
\includegraphics[width=9cm]{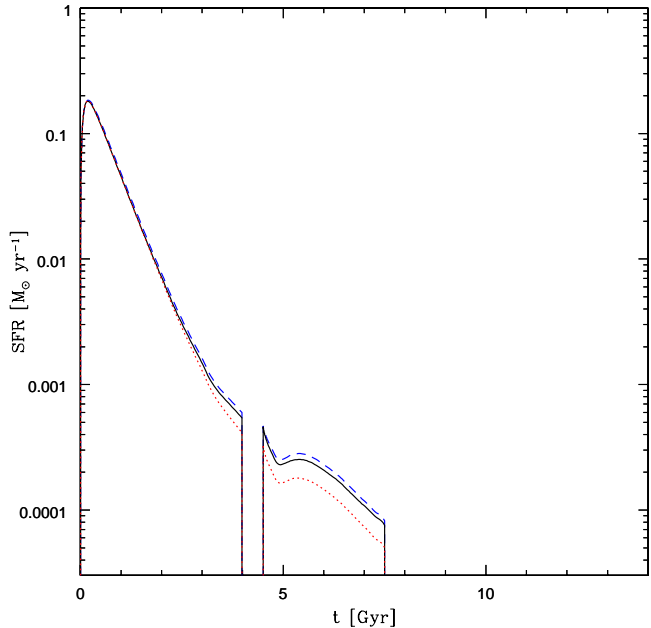}
\caption{In this figure, we compare the temporal evolution of the
  SFR as predicted by assuming the
  metallicity-dependent IGIMF of R14 (black solid line)
  with the same quantity predicted by assuming the \citet[dotted line
  in red]{salpeter1955} and \citet[blue dashed line]{chabrier2003}
  IMFs.
  The trend of the SFR traces that of the gas mass content within the galaxy. 
  In all the cases considered here, the
  SFR is always much lower than $1\,\text{M}_{\sun}$yr$^{-1}$. 
  Notice that the various curves almost
    overlap. This is due to the fact that the IMF has little
    effect on the global mass budget.}
     \label{sfr_sagittarius}
\end{figure}  

\begin{figure}
\includegraphics[width=9cm]{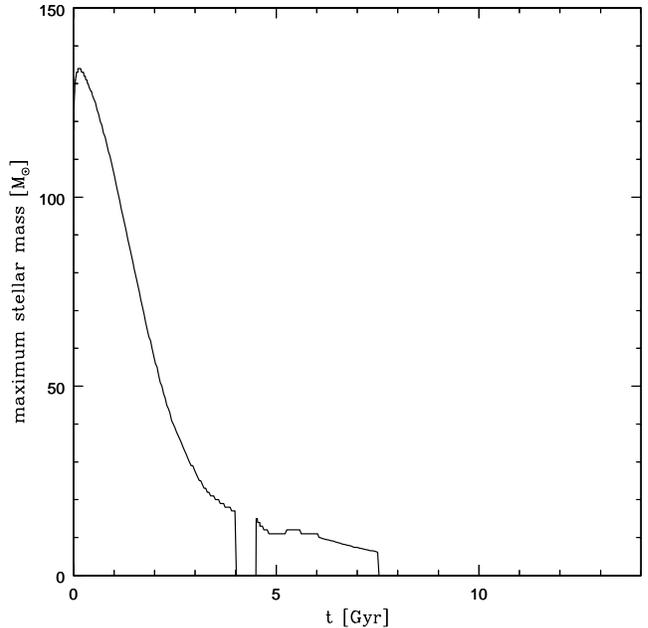}
\caption{Given the temporal evolution of the SFR and of the [Fe/H]
  abundance predicted by our chemical evolution model, this figure
  reports how the maximum stellar mass that can be formed at every
  time $t$ evolves as a function of the time itself, when assuming the
  metallicity-dependent IGIMF of R14. It is clear how
  much the truncation becomes important when the IGIMF is adopted in a
  detailed chemical evolution model of a galaxy with very low SFRs.  }
\label{m_max_sagittarius}
\end{figure} 
 
\section{The chemical evolution model}

\label{the_chemical_evolution_model} 

\subsection{The general model for dSphs}

In this work, we use a similar numerical code as described in
\citet{lanfranchi2004} and then adopted also in
\citet{lanfranchi2006}. The Sgr dSph is assumed to assemble
from infall of primordial gas into a pre-existing dark matter (DM)
halo, in a relatively short typical time-scale. The infall gas mass
has been set to $M_{\text{inf}}=5.0\times10^{8}\,\text{M}_{\sun}$ and the
infall time-scale has been set to $\tau_{\text{inf}}=0.5$ Gyr,
following the results of \citet{lanfranchi2006}. We assume
Sgr to have a massive and diffuse DM halo, with a mass
$M_{\text{DM}}=1.2\times10^{8}\,\text{M}_{\sun}$ \citep{walker2009} and
$S\equiv r_{\text{L}}/r_{\text{DM}}=0.1$, where $r_{\text{L}}=1550$ pc
\citep{walker2009} represents the effective radius of the baryonic
matter and $r_{\text{DM}}$ is the core radius of the DM halo. We need
the latter quantities in order to compute the potential well of the
gas and the time of the onset of the galactic wind, which is triggered
by the energy released into the ISM by the stellar winds and by the
core-collapse (Type II, Type Ib, Type Ic) and Type Ia SNe (see, for more
details, \citealt{bradamante1998} and \citealt{yin2011}). Once the
wind has started\footnote{In our model, the galactic wind develops
  when the thermal energy of the gas exceeds its binding energy to the
  galaxy.}, the intensity of the outflow rate is directly proportional
to the SFR.

\par The galaxy is modelled as a one-zone within which the mixing of
the gas is instantaneous and complete and the stellar lifetimes are
taken into account. We include the metallicity-dependent stellar
yields of \citet{karakas2010} for the low- and intermediate-mass
stars. For massive stars, we assume the He, C, N and O stellar yields at the various 
metallicities of 
\citet{meynet2002}, \citet{hirschi2005}, \citet{hirschi2007} and \citet{ekstrom2008} and, 
for heavier elements, the yields of \citet{kobayashi2006}.
Finally, we include the yields 
of \citet{iwamoto1999} for the Type Ia SNe. We assume the same stellar yields as 
\citet[see also \citealt{romano2010} for a detailed description]{matteucci2014}. 
It is worth noting that the yields of \citet{romano2010} have been selected 
because they are, at the present time, the best in order to reproduce the abundance patterns 
in the solar vicinity.

    \begin{figure}
\includegraphics[width=9cm]{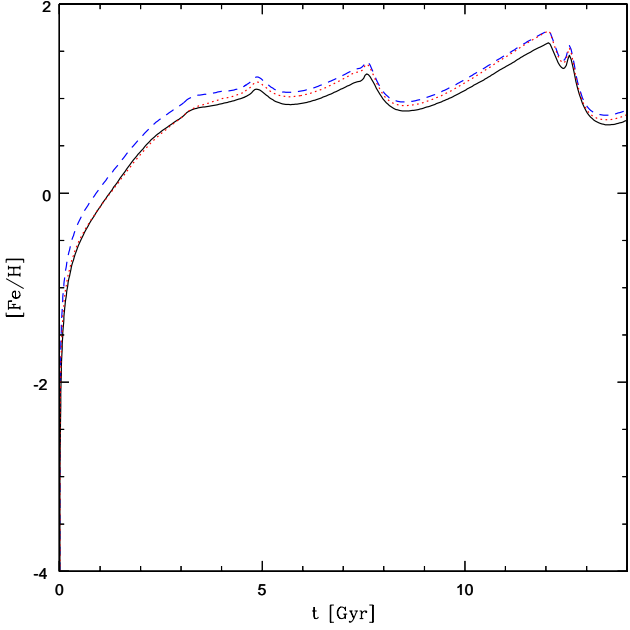}
\caption{In this figure, the predicted age--metallicity relation is shown. The $y$-axis
reports the [Fe/H] abundance of the galaxy ISM as predicted by our chemical evolution models, while the $x$-axis reports the time in
Gyr since the beginning of the galaxy evolution. The 
iron content within the ISM increases very steeply in the first billion years, then its trend flattens. 
The different IMFs considered in this work predict very similar age-metallicity relations. We remark on the fact 
that the [Fe/H] abundance in this figure refers to the abundance in the ISM; if one wants to see how many stars formed at a given time and therefore 
at a given [Fe/H] of the ISM, one should look at the stellar MDF (see Section \ref{results}). The lines correspond 
to the same IMFs as in Fig. \ref{sfr_sagittarius}.}
\label{age_metallicity}
\end{figure}     

\begin{figure}
\includegraphics[width=9cm]{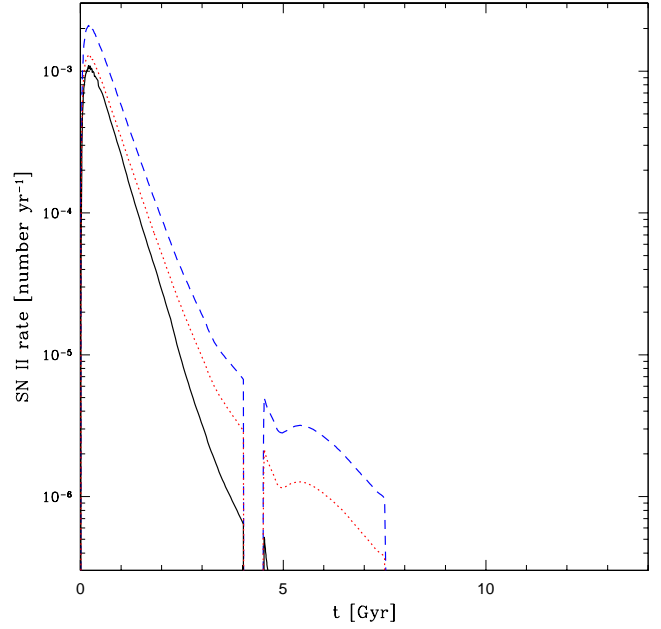}
\caption{This figure reports the core-collapse SN rate predicted as a
  function of the time. Because of the truncation of the IGIMF, the core-collapse SN
  rate predicted by the IGIMF is always lower than the one predicted
  by the classical IMFs. The lines correspond 
to the same IMFs as in Fig. \ref{sfr_sagittarius}.}
\label{ii}
\end{figure}   

\par We test in our model, separately, three different Eu nucleosynthetic yields: 
\begin{enumerate} 
 \item the yields of \citet[model 1, table 2]{cescutti2006}, in which the Eu is produced by core-collapse SNe, 
whose progenitors are massive stars with mass in the range $M=12$-$30\,\text{M}_{\sun}$;  
 \item the yields of \citet{ishimaru2004}, which can be found tabled in \citet[model 4, table 2]{cescutti2006}, 
where the Eu is produced by massive stars with mass in the range $M=8$-$10\,\text{M}_{\sun}$, exploding as core-collapse SNe; 
 \item the yield prescriptions of \citet{matteucci2014}, where we address the reader for details, for the Eu produced by NSMs.
 \footnote{Note that because of a typo mistake, the correct yield of Eu in \citet{matteucci2014} was $3.0\times10^{-6}\,M_{\odot}$ instead of 
 $3.0\times10^{-7}\,\text{M}_{\sun}$ (see \citealt{matteucci2015}).} 
 The first case we test is the model Mod3NS$^\prime$ (see fig. 6 of \citealt{matteucci2014}), where we assume that the Eu mass per NSM event 
 is $M_{\text{Eu,NSM}} = 3.0\times10^{-6}\,\text{M}_{\sun}$; the 
 progenitors of neutron stars lie in the range $9$-$50\,M_{\sun}$; the fraction of binary systems in this mass range becoming NSMs 
 is $\alpha_{\text{NSM}}=0.02$, and the time delay for NS coalescence is $\Delta t_{\text{NSM}}=1$ Myr (we check also $100$ Myr as 
 shown in Section \ref{results:alpha_mdf}). Moreover, we test also a model with 
 $M_{\text{Eu,NSM}} = 10^{-5}\,\text{M}_{\sun}$ per merger event, and the 
 other parameters being the same as Mod3NS$^\prime$. This value of the Eu yield is in agreement with the results 
 of recent calculations \citep{bauswein2014,just2014,wanajo2014}. 

\end{enumerate} 
  
\par If $M_{\text{g,}i}(t)$ is the gas mass in the form of an element
$i$ at the time $t$ within the ISM, the following basic equation
describes its temporal evolution in our chemical evolution model:
\begin{equation}
  \dot{M}_{\text{g,}i}=-\psi(t)X_{i}(t)+R_{i}(t)+(\dot{M}_{\text{g,}i})_{\text{inf}}-(\dot{M}_{\text{g,}i})_{\text{out}}. \label{eq:dsph}
\end{equation}
The quantity $ X_{i}(t)=M_{\text{g,}i}(t)/M_{\text{gas}}(t)$ is the
abundance by mass of the element $i$, with $\sum_{i}{X_{i}(t)}=1$ and
$M_{\text{gas}}(t)$ being the total gas mass of the galaxy at the time
$t$. The first term in the right hand side of equation (\ref{eq:dsph})
represents the rate of subtraction of the gas mass in the form of an
element $i$ because of the star formation activity, with the SFR
following the classical law of \citet{schmidt1959}: \begin{equation}
  \psi(t)=\Big(\frac{dM_{\text{gas}}}{dt}\Big)_{\text{SF}}=\nu M^{k}_{\text{gas}}, \label{eq:schmidt_law}
\end{equation} where $\nu$ is the so-called star formation
efficiency, expressed in [Gyr$^{-1}$], and $k=1$. $R_{i}(t)$ in equation (\ref{eq:dsph})
represents the ejected mass in the form of an element $i$ returned
per unit time by stars in advanced stages of their evolution. All
the prescriptions concerning the stellar yields and supernova
progenitor models are contained in this term.
\par The third term in equation (\ref{eq:dsph}),
$(\dot{M}_{\text{g,}i})_{\text{inf}}$, represents the rate of
accretion of the element $i$ during the infall event.  Since the gas
has initially a primordial chemical composition, we assume that
$X_{i\text{,inf}}=0$ for heavy elements. The infall event is assumed
to follow a decaying exponential law with $\tau_{\text{inf}}$ as
typical time-scale.
The last term in equation (\ref{eq:dsph}) represents the outflow rate
in the form of an element $i$, which is assumed to obey the following
law: \begin{equation}
  (\dot{M}_{\text{g,}i})_{\text{out}}=\omega_{i}\,\psi(t)=\omega_{i}\cdot(\nu\,M_{\text{gas}})=\lambda_{i}\,M_{\text{gas}}, \end{equation} 
  where $\lambda_{i}=\nu\cdot\omega_{i}$ is the so-called efficiency of the galactic wind for a given element $i$ (in units of Gyr$^{-1}$) which is 
the same here for all the chemical elements ($\lambda_{i}=\lambda$).

\subsection{The model for Sagittarius}

\par In this study we follow the results of \citet{lanfranchi2006},
which provide an estimate of the parameters of the chemical evolution
model for Sgr able to reproduce the observational data. They
found that Sgr should have been characterized by intermediate
values of the star formation efficiencies $\nu$, included between $1$
and $5$ Gyr$^{-1}$, and by intense galactic winds, with efficiencies $\lambda$
varying between $9$ and $13$ Gyr$^{-1}$.
\citet{lanfranchi2006} assumed a constant \citet{salpeter1955} IMF.

   \begin{figure}
\includegraphics[width=9cm]{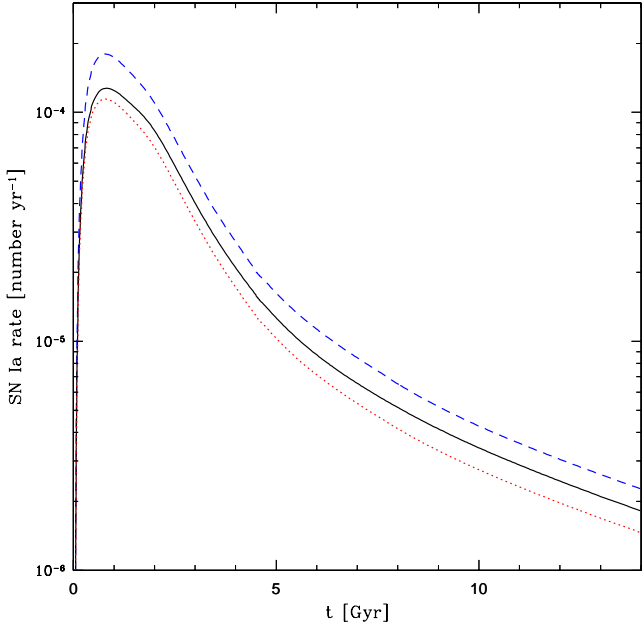}
\caption{In this figure, we compare the Type Ia SN rate as a function of the time. 
The integrated number of SNe Ia predicted with the \citet{chabrier2003} IMF turns out 
to be the largest one. The lines correspond 
to the same IMFs as in Fig. \ref{sfr_sagittarius}.}
\label{ia}
\end{figure}  

\begin{figure}
\includegraphics[width=9cm]{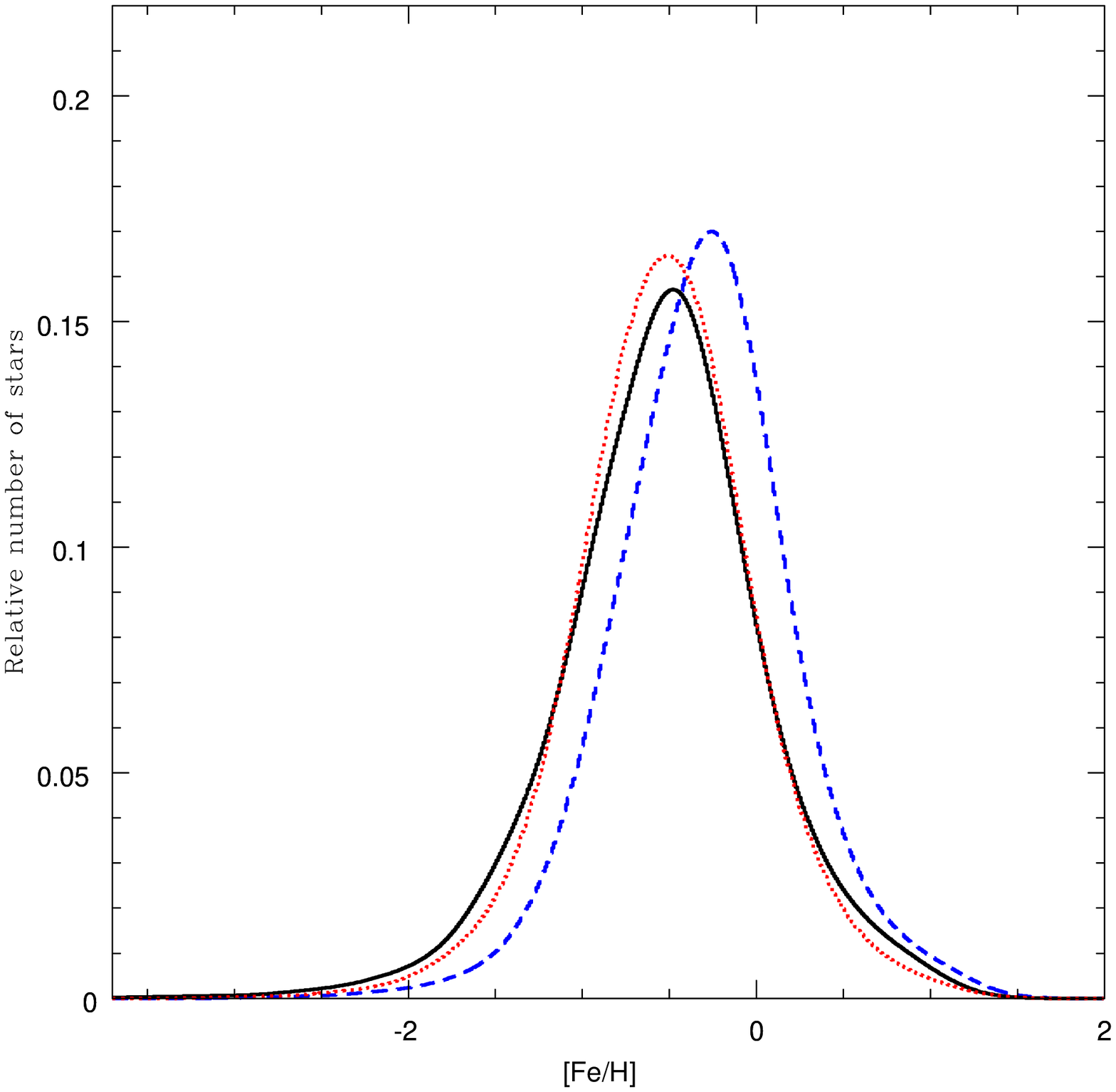}
\caption{In this figure, we compare the stellar MDF as predicted by
  the various models. The theoretical MDFs have been smoothed with a Gaussian
  function having $\sigma=0.2$. The IGIMF and the \citet{salpeter1955}
  IMF predict the MDF to peak at [Fe/H]=-0.48 dex and [Fe/H]=-0.51 dex, respectively, 
  in agreement with the mean value $\langle [Fe/H] \rangle = -0.5 \pm 0.2$ dex, measured
  by \citet{cole2001}. The \citet{chabrier2003} predicts the peak to
  occur at [Fe/H]=-0.26 dex. The lines correspond 
to the same IMFs as in Fig. \ref{sfr_sagittarius}.}
     \label{mdf}
\end{figure}  

\par In accordance to the observations, we assume that Sgr is
composed of two distinct stellar populations, one of old age $\geq10$
Gyr (the blue horizontal branch population discovered by
\citealt{monaco2003}) and one of intermediate age, which date back to
$8.0\pm1.5$ Gyr (the so-called Population A, studied by
\citealt{bellazzini2006}). So we adopt for the galaxy a star formation
history with two separate episodes, the first one occurring between
$0$ and $4$ Gyr since the beginning of the galaxy evolution, the
second one between $4.5$ and $7.5$ Gyr. Thus, according to these observational 
evidences, the star formation is set to zero outside those time intervals. 
During the star formation episodes, the SFR follows the Schmidt law 
(see equation \ref{eq:schmidt_law}).

  \begin{figure*}
\includegraphics[width=14cm]{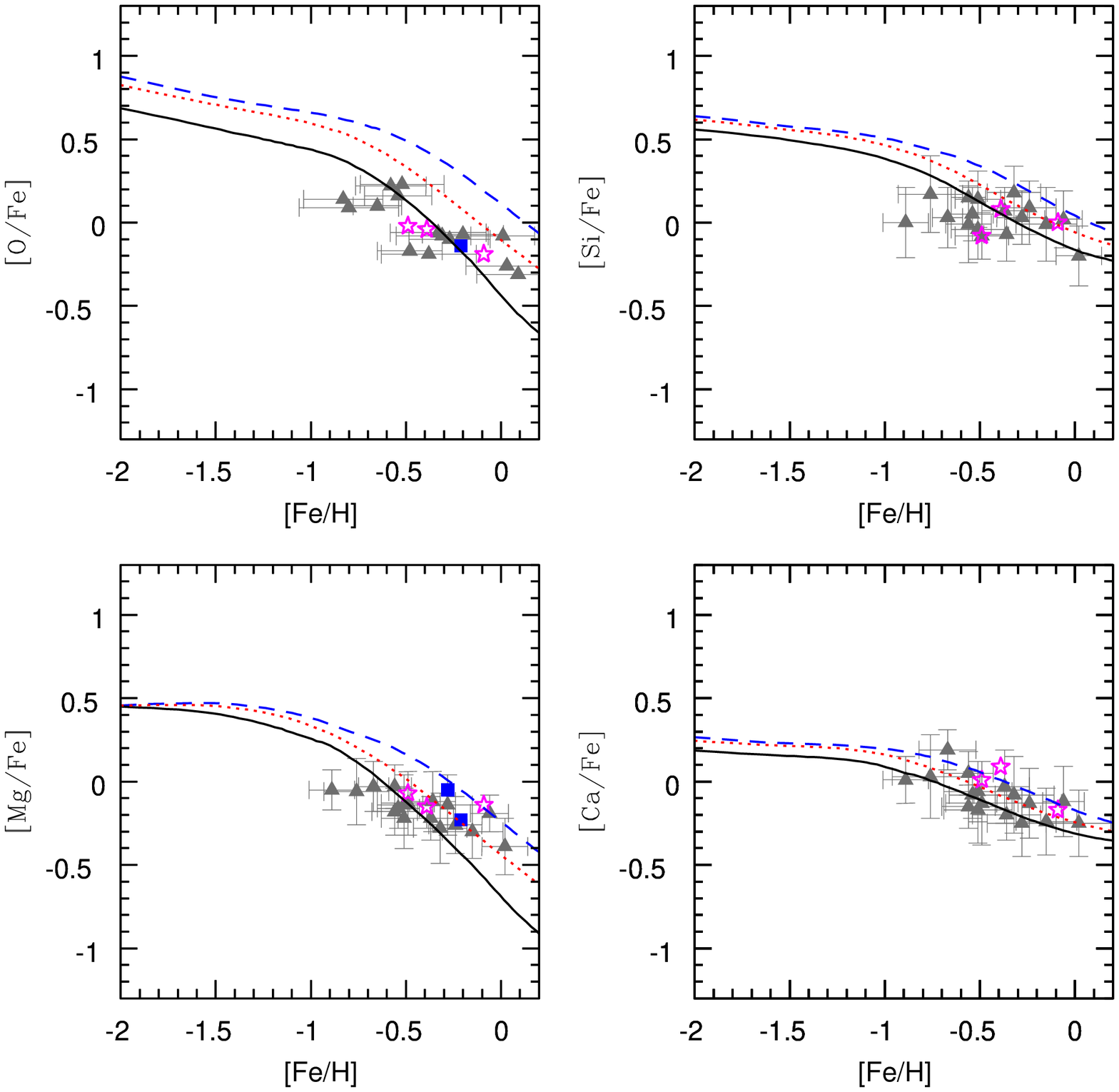}
\caption{In this figure, we report the [$\alpha$/Fe] versus [Fe/H]
  abundance ratio patterns as predicted by the IGIMF of R14 (black
  solid line) and by the \citet[dashed line in
  red]{salpeter1955} and \citet[blue dashed line]{chabrier2003}
  IMFs. The data are from \citet[blue squares]{bonifacio2000,bonifacio2004}, 
  \citet[grey triangles]{sbordone2007} and
  \citet[magenta stars]{mcwilliam2013}. The trend can be easily
  explained by means of the time-delay model
  \citep{matteucci1990,matteucci2001,lanfranchi2004} and by looking at
  the core-collapse and Type Ia SN rates in
  Fig. \ref{ii} and \ref{ia}, respectively. The lines correspond 
to the same IMFs as in Fig. \ref{sfr_sagittarius}.}
     \label{alpha}
\end{figure*}

\par In order to test what is the effect of the IGIMF in the framework
of a detailed chemical evolution model, we fix the following values
for the parameters of the model: $\nu=3$ Gyr$^{-1}$ and $\lambda=9$ Gyr$^{-1}$,
which are the best parameters found by \citet{lanfranchi2006}. We then
compare the results obtained by assuming the IGIMF with the ones
obtained by assuming the canonical \citet{salpeter1955} and
\citet{chabrier2003} IMFs. The \citet{salpeter1955} IMF is a
single-slope power law, which has the following form: \begin{equation}
  \phi(m)\propto m^{-2.35}\text{, for $0.1 \leq \frac{m}{\text{M}_{\sun}} < 100$,} \label{eq:salpeter}
\end{equation} whereas the \citet{chabrier2003} has a log-normal
distribution function for low-mass stars:
  \begin{equation}
\phi(m) \propto \left\{ \begin{array}{l l}
    \frac{1}{m} \exp( -\frac{(\log(m) - \log(0.079) )^{2}}{2\cdot0.69^{2}}) \text{, for $0.1 \leq \frac{m}{\text{M}_{\sun}} < 1$}\\
    m^{-2.3} \text{, for $1 \leq \frac{m}{\text{M}_{\sun}} < 100$.}
  \end{array} \right. \label{eq:chabrier}
\end{equation} 
In our chemical evolution model the minimum possible stellar mass is
$M_{\text{low}}=0.1\,\text{M}_{\sun}$, whereas the maximum possible stellar mass is
$M_{\text{up}}=100\,\text{M}_{\sun}$; so, if the maximum stellar mass of the IGIMF turns
out to be larger than $\text{M}_{\text{up}}$, we set it at the maximum
possible value of $100\,\text{M}_{\sun}$. The reason for that resides in the
fact that stars more massive than $100\,\text{M}_{\sun}$ have a negligible
effect in any IMF and it is difficult to find yields for them in the
literature.

  \begin{figure*}
\includegraphics[width=14cm]{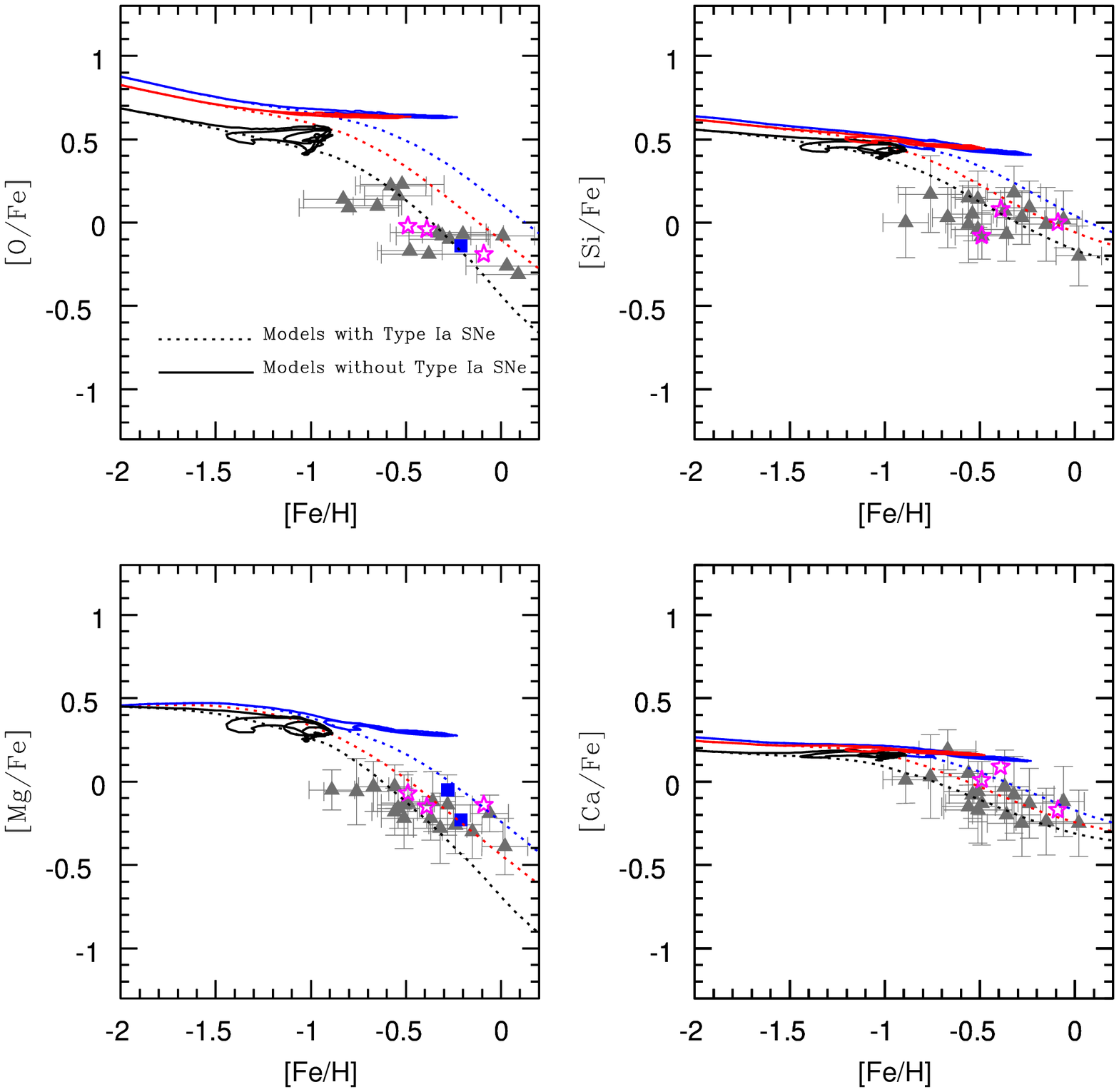}
\caption{In this figure, we study the effect on the predicted [$\alpha$/Fe] versus [Fe/H] 
abundance pattern of suppressing Type Ia SNe. The dotted lines correspond to the model with 
the inclusion of Type Ia SNe, while the solid lines represent the model when the contribution of 
Type Ia SNe has been suppressed. The dotted lines correspond 
to the same IMFs as in Fig. \ref{alpha}.}
     \label{alpha_noIa}
\end{figure*}

\begin{figure*}
\includegraphics[width=10cm]{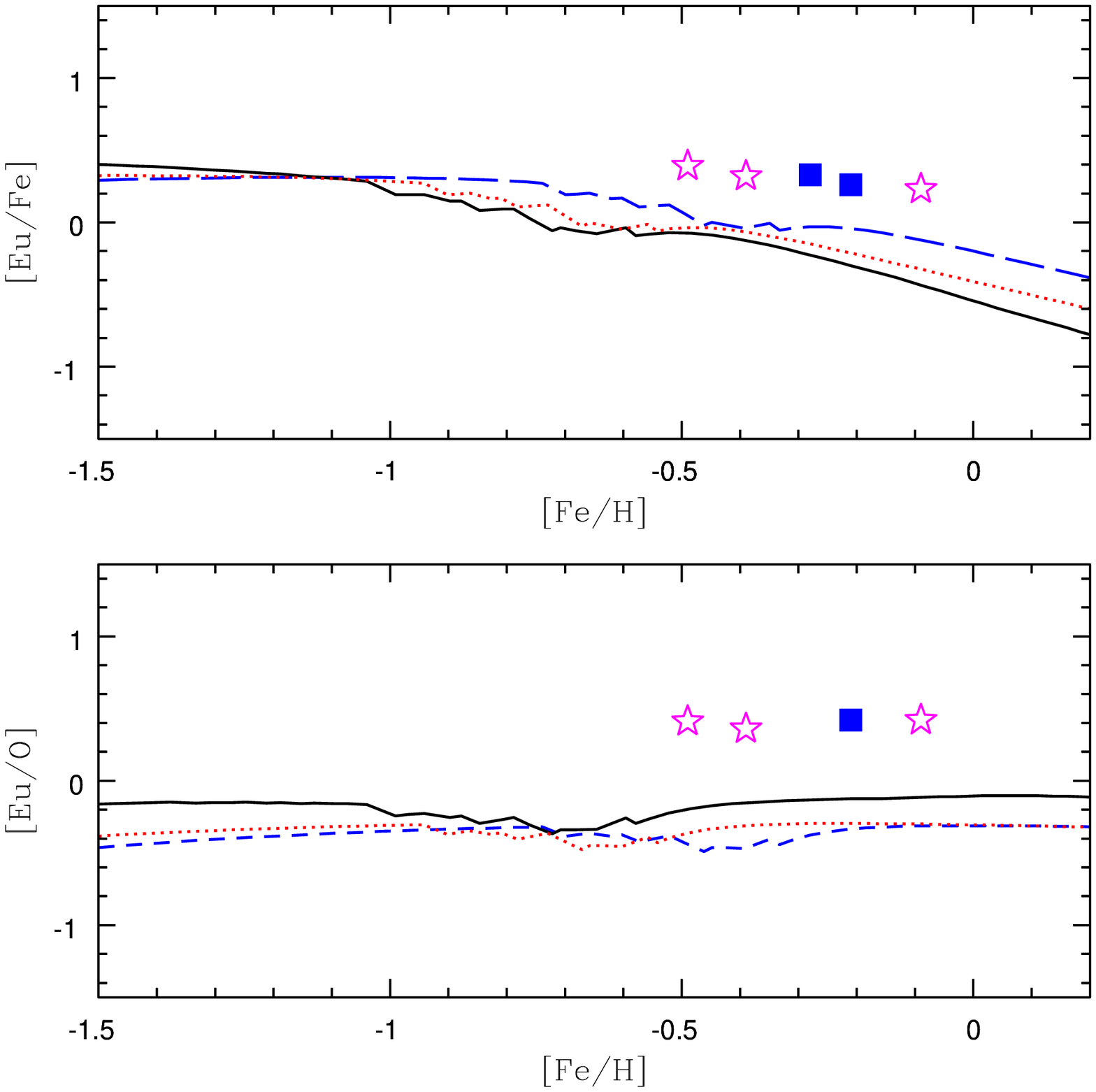}
\caption{In these figures, we compare the predictions of our models with different IMFs for the [Eu/Fe] and [Eu/O] versus [Fe/H] abundance patterns, 
when the yield of \citet{cescutti2006} are included. The latter assume 
the Eu to be produced by massive stars with mass in the range $M=12$-$30\,\text{M}_{\sun}$, which explode as core-collapse SNe.
None of the models with these yields is able to reproduce both the [Eu/Fe] and [Eu/O] abundance ratio patterns at the same time. 
The various lines correspond to the same 
IMFs as in Fig. \ref{sfr_sagittarius}.}
     \label{eu_cescutti}
\end{figure*}

\begin{figure*}
\includegraphics[width=10cmcm]{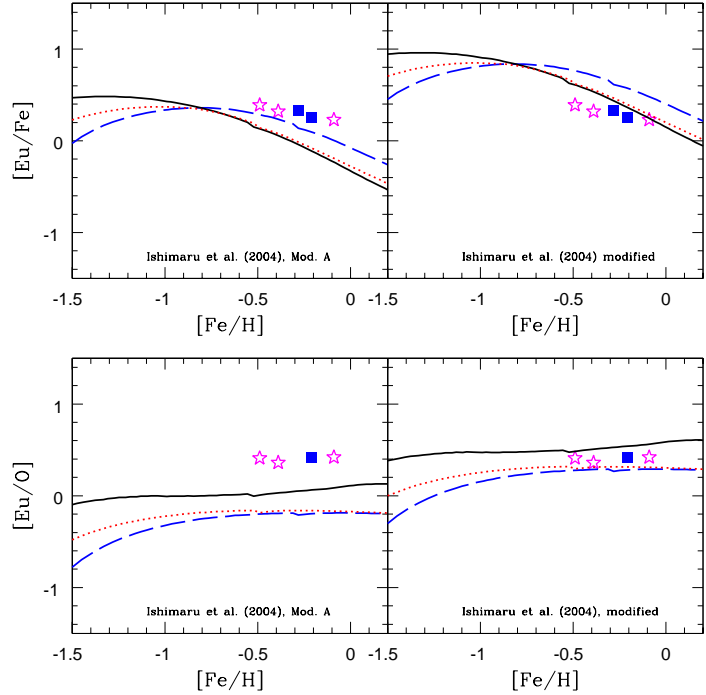}
\caption{In the top- and bottom-left figures, we compare the predictions of our models with different IMFs for the [Eu/Fe] and [Eu/O] versus [Fe/H] abundance patterns, 
when the yield of \citet{ishimaru2004} are included. The latter assume 
the Eu to be produced by stars with mass in the range $M=8$-$10\,\text{M}_{\sun}$, which explode as core-collapse SNe.
None of the models with these yields is able to reproduce both the [Eu/Fe] and [Eu/O] abundance ratio patterns at the same time. 
In the top- and bottom-right figures, we show the predictions of our models when the Eu yields of \citet{ishimaru2004} are multiplied by a factor of $3$; in this case, we 
can obtain a better results both for the [Eu/Fe] and the [Eu/O] abundance ratios, which can be reproduced by the model the IGIMF. 
The various lines correspond to the same 
IMFs as in Fig. \ref{sfr_sagittarius}.}
     \label{eu_ishimaru}
\end{figure*}

\begin{figure*}
\includegraphics[width=10cmcm]{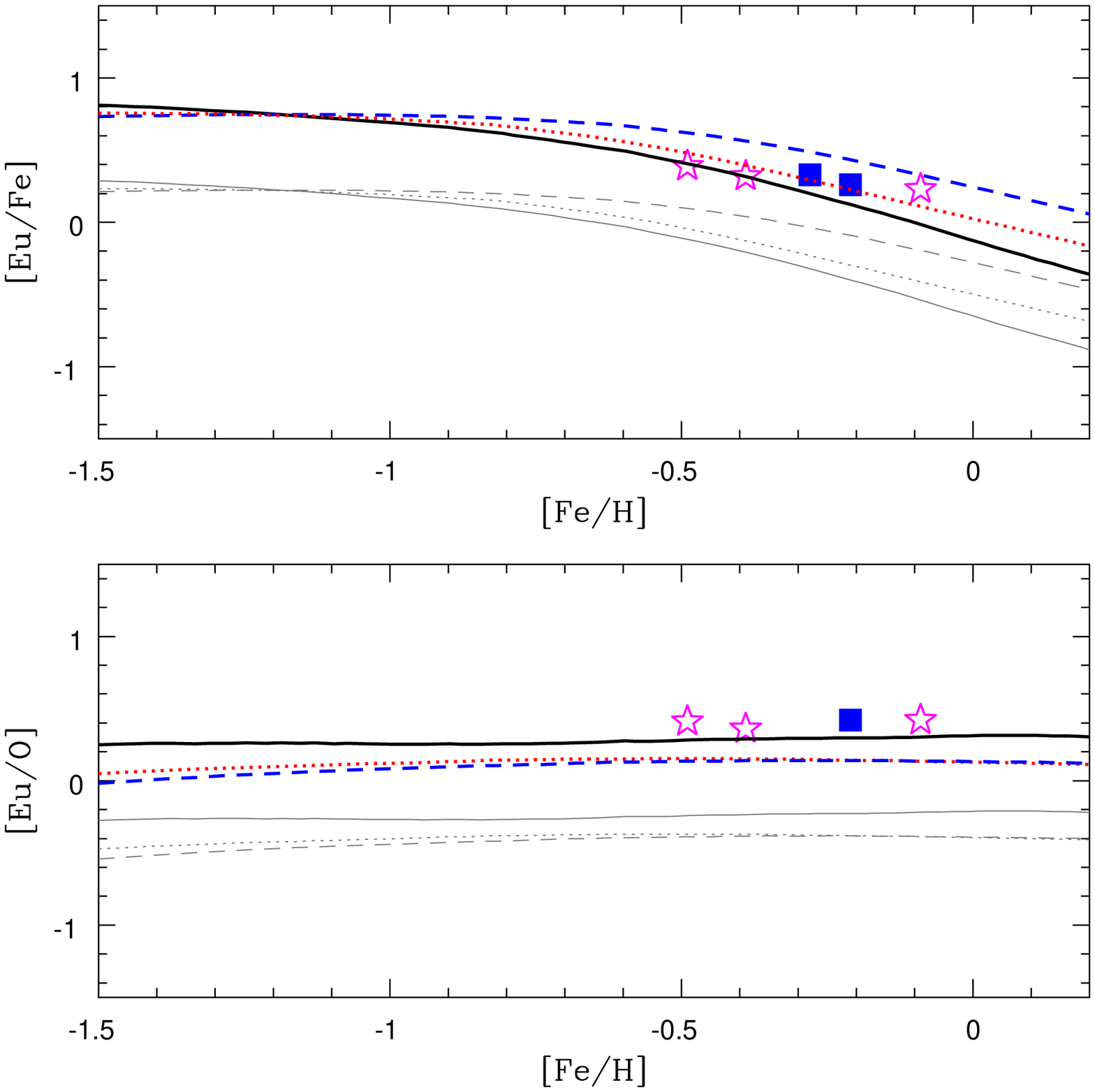}
\caption{In this figure, we compare the [Eu/Fe] and [Eu/O] as functions of [Fe/H] 
as predicted by the IGIMF and by the classical \citet{salpeter1955} and
\citet{chabrier2003} IMFs. The models which assume an Eu mass per NSM event $M_{\text{Eu,NSM}}=1.0\times10^{-5}\,\text{M}_{\sun}$ 
correspond to the thick coloured lines, whereas the models with $M_{\text{Eu,NSM}}=3.0\times10^{-6}\,\text{M}_{\sun}$ to the thin grey lines. 
The line crossing in the top figure around $\text{[Fe/H]} = -1.1$ is due to SNe Ia, which in the case of the IGIMF and Salpeter IMF 
start to explode when the [Fe/H] abundance of the ISM is lower than in the case with the Chabrier IMF. 
The various lines within each set correspond to the same IMFs as in Fig. \ref{sfr_sagittarius}. } 
     \label{europium}
\end{figure*}

\section{Results} \label{results}

This work is based on the chemical evolution model described in
Section \ref{the_chemical_evolution_model} and our aim is to
investigate the effect of three different IMFs: the canonical
\citet{salpeter1955}, the \citet{chabrier2003} IMF and the
metallicity-dependent IGIMF of R14.
\par The galaxy is always predicted to 
possess at the present time a very small total amount of gas. 
In particular, the total H\textsc{i} mass results 
$M_{\text{HI}}\approx1.8\times10^{4}\,\text{M}_{\sun}$ with the \citet{chabrier2003} IMF, 
$M_{\text{HI}}\approx1.3\times10^{4}\,\text{M}_{\sun}$  with the \citet{salpeter1955} IMF 
and $M_{\text{HI}}\approx1.8\times10^{4}\,\text{M}_{\sun}$ with the IGIMF. 
All the latter quantities are almost in agreement with the upper 
limit of the total H\textsc{i} mass derived by \citet{koribalski1994}, which is 
$M_{\text{HI,obs}}\sim10^{4}\,\text{M}_{\sun}$.
\par The model with the IGIMF predicts the largest final total stellar mass
for the galaxy, which is $M_{\star,\text{fin}}\approx1.1\times10^{8}\,\text{M}_{\sun}$. 
In fact, the model with the \citet{salpeter1955} and \citet{chabrier2003} 
IMFs predict $M_{\star,\text{fin}}\approx7.9\times10^{7}\,\text{M}_{\sun}$ and
$M_{\star,\text{fin}}\approx5.2\times10^{7}\,\text{M}_{\sun}$, respectively, which 
have the same order of magnitude of the observed total stellar mass 
$M_{\star}\sim2.1\times10^{7}\,\text{M}_{\sun}$ (see \citealt{mcconnachie2012} and 
references therein). In this study, we have neglected the fact that Sgr has 
lost many stars after its SF ceased; this may explain why the actual 
stellar mass predicted by our chemical evolution models is larger than the present-day 
observed mass.

\par The model
with the IGIMF predicts the galactic wind to develop for the first
time at $t_{\text{GW}}=30$ Myr while the model with the
\citet{salpeter1955} and the \citet{chabrier2003} IMFs predict the
onset of the galactic wind at $t_{\text{GW}}=25$ and $20$ Myr,
respectively. Since we assume the SFR to proceed since the beginning,
the retarded onset of the galactic wind with the IGIMF is likely due
to the strong truncation of IGIMF itself, which inhibits the formation
of very massive stars, the ones having the shortest typical lifetimes
and exploding as core-collapse supernovae. This can be confirmed by
the intensity of the SFRs under play; if they are
$\leq1\,\text{M}_{\sun}$yr$^{-1}$, then the truncation is important. By
looking at Fig. \ref{sfr_sagittarius}, where the predicted trend of
the SFR is plotted as a function of time, it turns out that the predicted
SFRs in Sgr are always much lower than
$1\,\text{M}_{\sun}$yr$^{-1}$. The temporal evolution of the maximum
stellar mass that can be formed during the star formation activity, in
the case of the IGIMF, is shown in Fig. \ref{m_max_sagittarius}. The
steep increasing trend of $m_{\text{max}}$ at the beginning is due to the
rapid increase of the SFR during the initial infall event. Then,
$m_{\text{max}}$ decreases because of the declining SFR. The [Fe/H]
dependence is crucial during the initial infall event, when the
[Fe/H] abundance rapidly increases, counterbalancing the
SFR dependence and preventing the IGIMF to reach masses very close to
the empirical limit of $150\,\text{M}_{\sun}$. In fact, the bulk of chemical
enrichment in this galaxy occurs in the first Gyr of its evolution;
then, the age--metallicity relation becomes much more shallow and the
main role is played by the SFR, which decreases very steeply. In Fig. 
\ref{age_metallicity}, we report the age--metallicity relations predicted 
by assuming the different IMFs. The small fluctuations visible in Fig. 
\ref{age_metallicity} are due to the bursting mode of star formation and the 

\par The core-collapse SN rate as a function of time is
shown in Fig. \ref{ii}. The
progenitors of core-collapse SNe are assumed to be massive stars with mass $M>8\,\text{M}_{\sun}$,
which have very short typical lifetimes since star formation, in the
range $1\,\text{Myr}<\tau_{M}<35\,\text{Myr}$.  As one can see from the figure,
the IGIMF predicts numbers of core-collapse SNe per unit time which are always lower 
than the ones predicted by the 
classical IMFs. On the other hand, among the IMFs here considered, the
highest number of stars over the entire high-mass range originate when
assuming the \citet{chabrier2003} IMF (see also \citealt{romano2005}). In fact, the
\citet{chabrier2003} IMF for $M\geq1\;\text{M}_{\sun}$ has a slope
$\alpha_{\text{Chab.}}=-2.3$ (see equation \ref{eq:chabrier}), which
is flatter than the \citet{salpeter1955} one ($\alpha_{\text{Sal.}}=-2.35$, 
see equation \ref{eq:salpeter}). 
For this reason, the core-collapse SN rate with the \citet{chabrier2003}
exceeds the other.

\par In Fig. \ref{ia}, we report the predicted Type Ia SN rate as a
function of time. We assume Type Ia SNe to originate from white dwarfs in
binary systems exploding by C-deflagration. We adopt the so-called 
\textit{single-degenerate model}, with the same 
prescriptions of \citet{matteucci2001}. According to this particular 
progenitor model, a degenerate C--O 
white dwarf (the primary, initially more massive, star) accretes 
material from a red giant or main-sequence companion (the secondary, 
initially less massive, star); in summary, when the white dwarf 
reaches the Chandrasekhar mass, the C-deflagration occurs and the 
white dwarf explodes as a Type Ia supernova (for more details, see 
\citealt{matteucci2001book}). Depending primarily on the mass of the 
secondary star, which is the clock for the explosion, 
Type Ia SNe can explode over a very large interval of 
time-scales since the star formation, which can vary between $35$ Myr 
and the age of the Universe. 
By looking at Fig. \ref{ia}, the Type Ia SN
rate with the \citet{chabrier2003} IMF dominates over the other two;
in fact, the \citet{chabrier2003} IMF predicts also a higher number of
low- and intermediate-mass stars with $M>1\,M_{\sun}$
\citep{romano2005}.

\subsection{MDF and chemical abundances} \label{results:alpha_mdf}

The stellar metallicity distribution function (MDF) predicted by the
model with the IGIMF has the peak at $\text{[Fe/H]=}-0.48$ dex, close to the
position of the peak predicted by the model with the
\citet{salpeter1955} IMF, which is at $\text{[Fe/H]}=-0.51$ dex. 
Conversely, the [Fe/H] peak of the MDF with the \citet{chabrier2003} 
IMF occurs at $\text{[Fe/H]}=-0.26$ dex, which is much higher than the other two values. 
This can be seen in Fig.
\ref{mdf} and it is due to the fact that the integrated number of Type
Ia and core-collapse SNe with the \citet{chabrier2003} IMF is much
higher than that predicted when assuming the IGIMF or the
\citet{salpeter1955} IMF (see Figs \ref{ii} and \ref{ia}). So, on
average, fixing all the other parameters of the model, a quite enhanced
iron pollution from SNe is expected when adopting the
\citet{chabrier2003} IMF.
Finally, the IGIMF and the
\citet{salpeter1955} IMF predict a [Fe/H] abundance for the peak of
the stellar MDF which is in agreement with the mean value $\langle
\text{[Fe/H]} \rangle = -0.5 \pm 0.2$ dex, measured by \citet{cole2001} for
the Sgr main stellar population.

\par In Fig. \ref{alpha}, we compare the predicted [$\alpha$/Fe] abundance ratios 
as a function of the [Fe/H] abundances with the observational data of 
\citet{bonifacio2000,bonifacio2004}, 
\citet{sbordone2007} and \citet{mcwilliam2013}. We remind the reader
that Type Ia SNe enrich the ISM mainly with iron (almost $2/3$ of the
total content) and iron-peak elements, whereas the $\alpha$-elements
are mainly produced by core-collapse SNe, which also provide some
iron, typically $\sim1/3$ of the total. However, some
$\alpha$-elements, such as the calcium and the silicon, are also
synthesized by Type Ia SNe, although in smaller quantities than those
coming from core-collapse SNe. We have also to remark the fact that
the fraction of the total iron content coming from Type II and Type Ia SNe 
depends on the assumed IMF and the aforementioned proportions have been
calculated by assuming Salpeter-like IMFs (see for more details, 
\citealt{matteucci2014_solo}). 
\par By looking at Fig. \ref{alpha}, the
overall trend predicted by assuming the three different IMFs is quite
similar and it can be easily explained by the so-called
\textit{time-delay model}
\citep{matteucci1990,matteucci2001,lanfranchi2004}: the decrease of
[$\alpha$/Fe] at very low [Fe/H] is due to the very low SFR under
play, which causes Type Ia SNe to be important in the iron pollution
when the ISM was not yet heavily enriched with iron by core-collapse
SNe; then, the further steepening of [$\alpha$/Fe] is due to the strong 
outflow rate. The fundamental role played by the so-called time-delay model 
in shaping the trend of the [$\alpha$/Fe] ratios, as a function of [Fe/H], is 
shown in Fig. \ref{alpha_noIa}, 
where we study the effect of suppressing Type Ia SNe in the chemical enrichment process.  
As one can see from the figure, if there are no Type Ia SNe, which are the most 
important Fe producers in galaxies, a truncated IMF such as the IGIMF would never be able to 
reproduce the data by itself. Furthermore, only by means of Type Ia SNe the galaxy can reach the observed [Fe/H] 
abundances. In fact, the predicted trend reflects only the contribution of core-collapse 
SNe to Fe. It is only the contribution of Type Ia SNe 
that can explain the decrease of [$\alpha$/Fe] ratios and produce the right amount of Fe.

\begin{figure*}
\includegraphics[width=10cm]{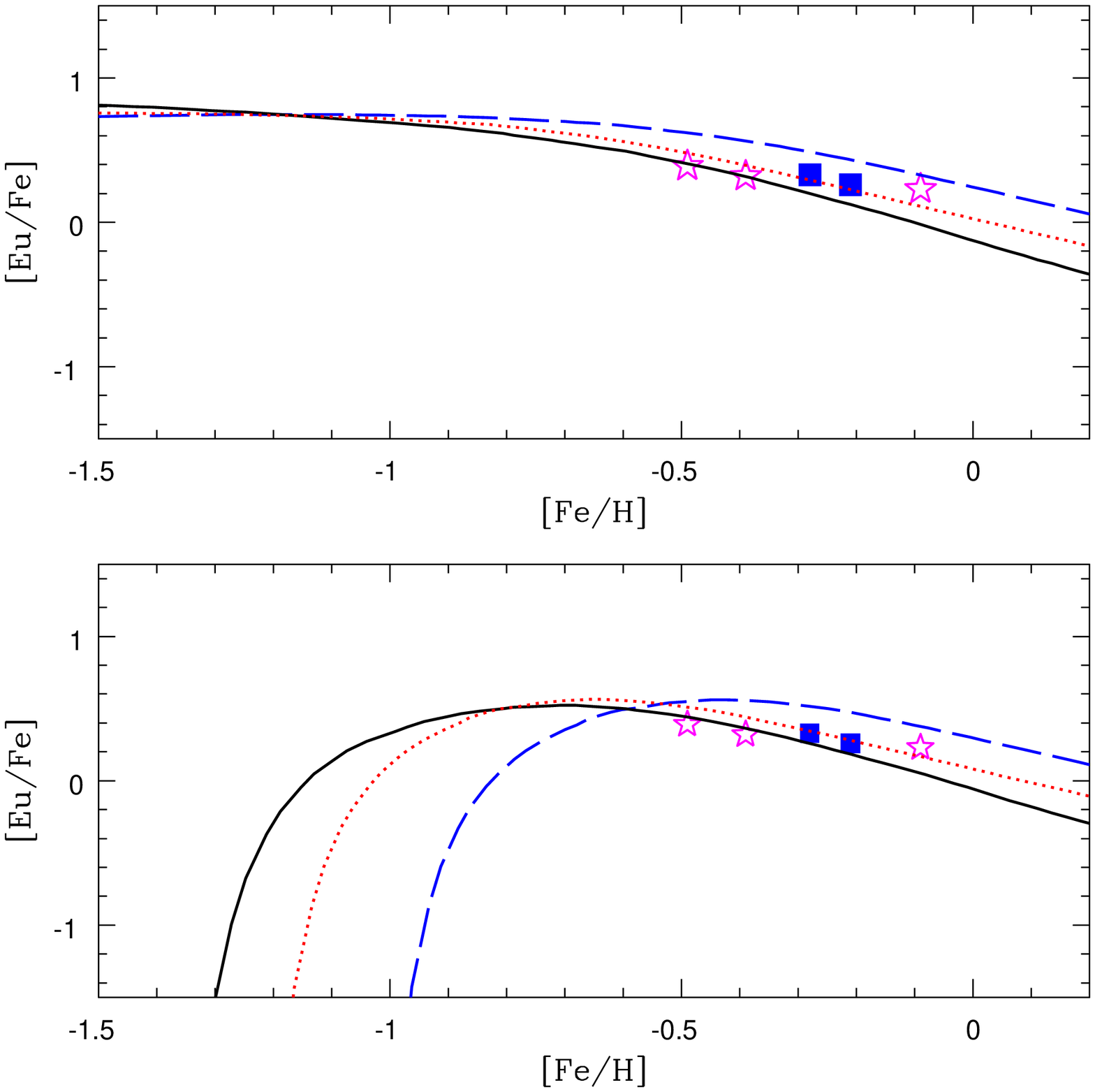}
\caption{In this figure, we show the effect on the [Eu/Fe] versus [Fe/H] relations of varying the delay time for the coalescence of the NS close binary system from 
$\Delta t_{\text{NSM}}=1$ Myr (top figure) to $100$ Myr (bottom figure). The various lines correspond to the same 
IMFs as in Fig. \ref{sfr_sagittarius} and all the models assume a mass of Eu per NSM event $M_{\text{Eu,NSM}}=1.0\times10^{-5}\,\text{M}_{\sun}$. } 
     \label{europium_tau}
\end{figure*}

\begin{figure}
\includegraphics[width=9cm]{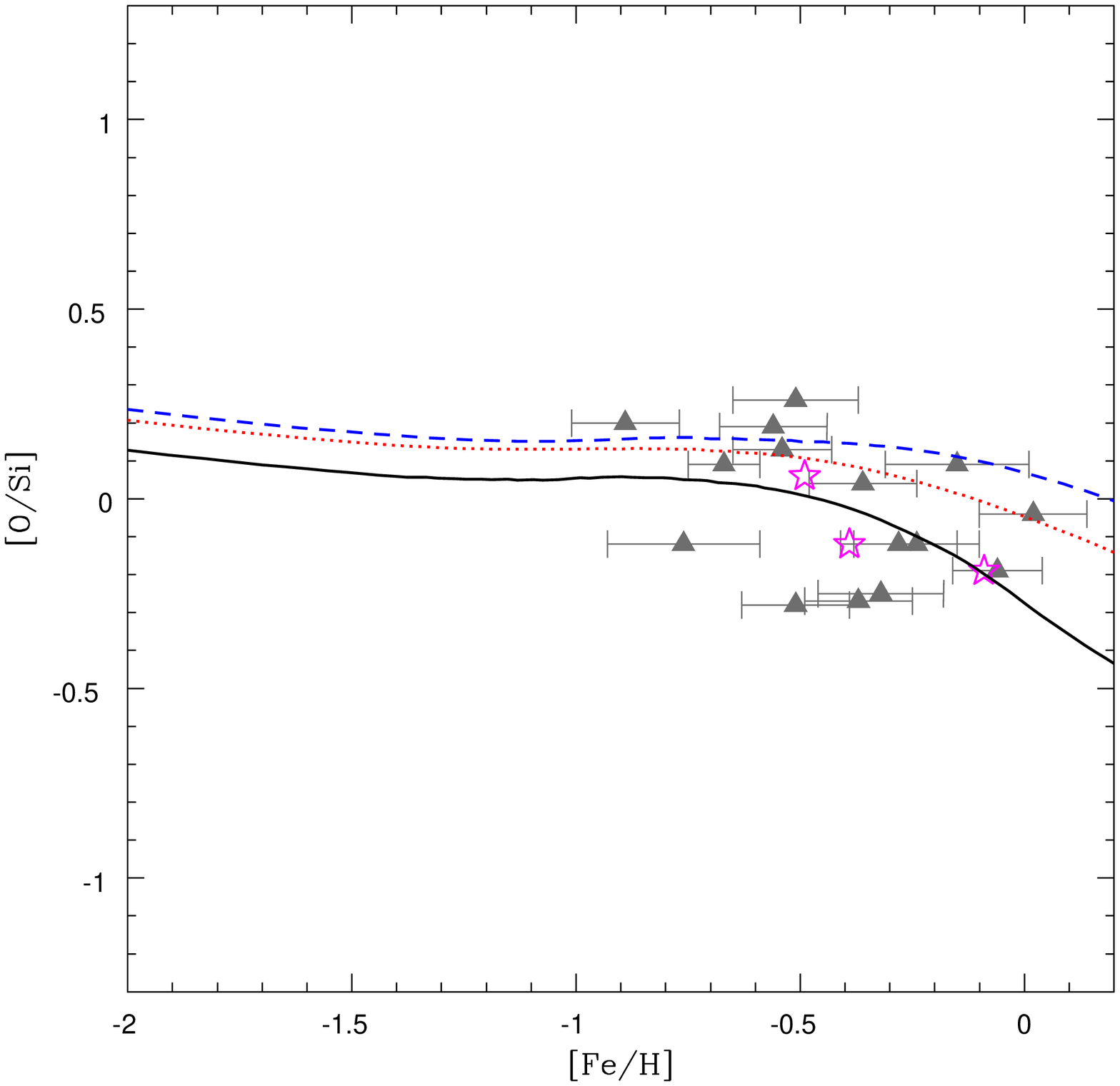}
\caption{In this figure, we report the predictions 
of our models for the [O/Si] ratios (hydrostatic over explosive $\alpha$-element ratios) 
as functions of [Fe/H], in order to ascertain if the data suggest a truncated IMF. The various lines correspond to the same 
IMFs as in Fig. \ref{sfr_sagittarius}.}
     \label{abundance_ratios}
\end{figure} 

\par The position of the knee in the [$\alpha$/Fe] ratios as a function of [Fe/H] varies 
from galaxy to galaxy and it primarily depends upon the total mass of the galaxy, where the dSphs 
with larger mass exhibit the knee preferentially at higher [Fe/H]. We explain this fact by assuming higher 
efficiency of SF for dwarf galaxies of larger total mass, with the low mass ultrafaint dwarf spheroidals  
needing the lowest SF efficiencies \citep{salvadori2009,vincenzo2014}.

\par By looking at Fig. \ref{alpha}, for a given value of [Fe/H], 
the model with the IGIMF predicts the lowest [$\alpha$/Fe] abundances while 
the highest [$\alpha$/Fe] ratios are reached when assuming the \citet{chabrier2003} IMF. 
In order to obtain the same high values of [$\alpha$/Fe] in a model with the IGIMF,
one should slightly increase the star formation efficiency. This can be also explained 
by looking at Fig. \ref{ii}, from which one can conclude that the \citet{chabrier2003} IMF 
predicts the highest core-collapse SN rates, whereas the IGIMF the lowest ones, over the 
entire galaxy lifetime.
So in conclusion, given a particular value of the 
galaxy gas mass fraction $\mu = M_{\text{gas}}/M_{\text{tot}}$, the \citet{chabrier2003} 
IMF predicts the 
highest metal content in the galaxy while the IGIMF predicts the lowest one. 

\par In Fig. \ref{eu_cescutti}, we show the predictions of our models with the Eu yields of \citet{cescutti2006} for
the [Eu/Fe] versus [Fe/H] abundance ratio patterns. We remind the reader that in this case the Eu is assumed to be produced by 
core-collapse SNe in the range $M=12$-$30\,\text{M}_{\sun}$ and the stellar yields were determined ad hoc to reproduce the 
observational trend observed in the MW stars. We find that neither the IGIMF 
nor the classical IMFs are able to reproduce the observed data set when adopting those yields. 

\begin{figure*}
\includegraphics[width=14cm]{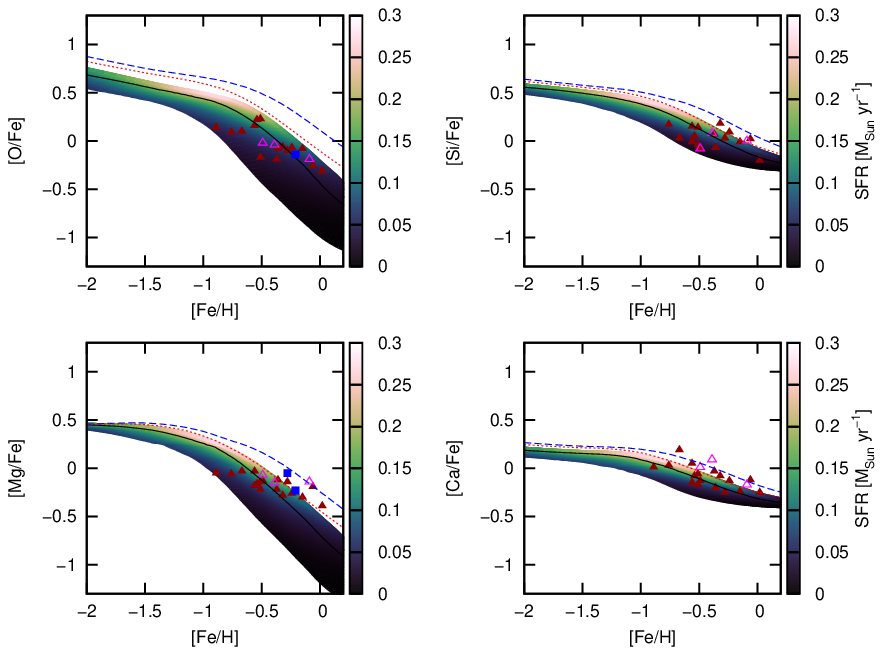}
\caption{In this figure, we show what is the effect of varying the $\nu$ parameter 
in the [$\alpha$/Fe] versus [Fe/H] abundance pattern when the IGIMF is assumed.
The colour-coded curves have been obtained by varying the SF efficiencies in 
the range $\nu=1-5$ Gyr$^{-1}$, with the wind parameter fixed at the value $\lambda=9$ Gyr$^{-1}$.
The colour-coding represents the SFR expressed in units of $\text{M}_{\sun}$ yr$^{-1}$ and 
the model with the lowest SF efficiency ($\nu=1$ Gyr$^{-1}$) corresponds to
the lowest edge of the plot, while the model with the highest SF efficiency 
($\nu=5$ Gyr$^{-1}$) corresponds to the highest edge.
The black solid line, the dotted line in red and the dashed line in blue correspond to the reference 
models with $\nu=3$ Gyr$^{-1}$ and $\lambda=9$ Gyr$^{-1}$ with the IGIMF, the \citet{salpeter1955} and the 
\citet{chabrier2003} IMFs, respectively, as shown in Fig. \ref{alpha}; also the data are the same as in Fig. \ref{alpha}.}
     \label{var_nu}
\end{figure*}

\par \citet{wanajo2003} and \citet{mcwilliam2013} proposed 
that Eu could be produced by core-collapse SNe 
whose progenitors are less massive than the stars more important in oxygen production, 
which have masses $\ga30\,M_{\sun}$. 
We tested this scenario in our chemical evolution model of the Sgr dwarf. Therefore, 
we have included in our models the Eu yields by \citet{ishimaru2004} from core-collapse SNe in the range $8$-$10\,\text{M}_{\sun}$. According 
to these yields, the Eu is produced as an r-process element, with 
$X_{\text{Eu}}^{\text{new}}=3.1\times10^{-7}\,\text{M}_{\sun}/M_{\star}$ being the fraction of Eu ejected by a star of mass $M_{\star}$. 
The results of the chemical evolution models which assume the yields of \citet{ishimaru2004} can be seen in Fig. 
\ref{eu_ishimaru}, where we show the [Eu/Fe] abundance ratio as a function of the [Fe/H] abundances, as predicted by our models 
with different IMFs. In that figure, we show also the results of models with the \citet{ishimaru2004} yields artificially increased 
by a factor of $3$. We find that the models with the original \citet{ishimaru2004} yields are not able to reproduce [Eu/Fe] and [Eu/O] at the same time. 
In fact, while the [Eu/Fe] ratios can be better explained by the models with the \citet{chabrier2003} IMF, because of the higher weight of the $8$-$10\,\text{M}_{\sun}$ 
stars in this IMF, the high [Eu/O] ratios cannot be reproduced even by the model with the IGIMF, which predicts a lack of O. By increasing the \citet{ishimaru2004} 
yields, we obtain a better result and, in principle, we could explain in this way the observed Eu abundances in this galaxy. 

\par Since the r-process nucleosynthesis is still debated 
in the literature, we also tested the case in which NSM events are the main responsible for the production of Eu in galaxies, a hypothesis which has received 
a large interest recently (e.g. \citealt{mennekens2014,matteucci2014,vandevoort2015,wehmeyer2015}).
In Fig. \ref{europium}, the [Eu/Fe] and [Eu/O] ratios as predicted by our models with $M_{\text{Eu,NSM}}=1.0\times10^{-5}\,\text{M}_{\sun}$ 
(thick lines) are compared with the ones predicted by our models assuming $M_{\text{Eu,NSM}}=3.0\times10^{-6}\,\text{M}_{\sun}$ (thin lines), as in \citet{matteucci2014}. 
In both cases, the truncation of the IGIMF strongly affects the predicted [Eu/O] ratios, which are always higher than the [Eu/O] ratios predicted with the 
standard IMFs. 
Our models with $M_{\text{Eu,NSM}}=3.0\times10^{-6}\,\text{M}_{\sun}$ are still not able to reproduce the abundance ratios in the Sgr dwarf. 
The choice of the  $M_{\text{Eu,NSM}}=3.0\times10^{-6}\,\text{M}_{\sun}$ derives from the best value quoted by Matteucci et al. (2014) to reproduce Eu in the solar vicinity: 
however, the yields of Eu per event can be as high as $M_{\text{Eu,NSM}}=1.0\times10^{-5}\,\text{M}_{\sun}$, in agreement with the upper limit of \citet{korobkin2012} and 
with current calculations adopting more recent nuclear data (e.g., \citealt{wanajo2014}). With this value we can clearly better reproduce 
both the [Eu/Fe] and the [Eu/O] abundance ratios observed in this galaxy. 
Because of the mentioned still existing nucleosynthesis uncertainties and the small number of observations available for dSph galaxies and Sgr galaxy in particular, 
our results can only safely demonstrate that the idea of \citet{mcwilliam2013} is correct and that the [Eu/O] ratio can be a possible diagnostic in future observations and 
studies of chemical evolution. In this context, we do not wish to explore all the possible combinations for Eu production sites, as in \citet{matteucci2014} 
where models including both core-collapse and NSMs were considered. The paucity of data for Sgr, in fact,  prevents us from drawing any conclusions on Eu produced in stars with 
mass as large as $50\,\text{M}_{\sun}$, leaving their chemical signature at low metallicities. For the same reason, we cannot safely conclude anything about the time delay for the coalescence 
of neutron stars. To illustrate that, in Fig. \ref{europium_tau}, we show what is the effect on the predicted [Eu/Fe] versus [Fe/H] relations of varying the delay time for coalescence from 
 $\Delta t_{\text{NSM}}=1$ Myr (top figure) to $\Delta t_{\text{NSM}}=100$ Myr (bottom figure). We remark on the fact that these are extreme cases, 
 given the uncertainty still present in the delay time for NSMs 
 (see \citealt{dominik2012,vandevoort2015}). 

\par In Fig. \ref{abundance_ratios}, we show also the abundance patterns 
of [O/Si]. Following the suggestions of \citet{mcwilliam2013}, 
the truncation of the IMF can leave a signature in the hydrostatic over explosive $\alpha$-element abundance ratios. 
The Si is an explosive $\alpha$-elements and its stellar yields are not affected by the truncation as much as 
those of oxygen. By looking at the Figure, the \citet{mcwilliam2013} data for [O/Si] are well reproduced with the 
IGIMF, supporting the idea that a truncated IMF should be preferred in this galaxy. However, the data are still  
uncertain and prevent us from drawing firm conclusions.

\par It is interesting to note that the dispersion in the [r-process/Fe] abundance ratios observed in the extreme metal-poor halo stars 
suggests that the frequency of r-process producers, per SN event, must be $\sim5$ per cent \citep{mcwilliam1995,fields2002}. 
This could be considered as a support to the idea of NSMs as Eu producers, since NSM binaries are a small fraction of the number of core-collapse SN events 
(we assume $\alpha_{\text{NSM}}=0.02$, as in \citealt{matteucci2014}).

\subsection{Exploring the parameter space}

\label{expl}

In Fig. \ref{var_nu}, we explore the effect of changing the model parameter
$\nu$ in the [$\alpha$/Fe] versus [Fe/H] abundance ratio patterns, when
the metallicity-dependent IGIMF of R14 is assumed. 
The third dimension (colour-coding) in the figure corresponds to the SFRs under play and the 
parameter space that we explore is the one provided by \citet{lanfranchi2006}, with the SF 
efficiencies continuously varying in the range $\nu=1$-$5$ Gyr$^{-1}$. 
We remark on the fact that \citet{lanfranchi2006} assumed a \citet{salpeter1955} IMF.
\par By looking at Fig. \ref{var_nu}, by increasing the SF efficiency, it allows us to reach higher [$\alpha$/Fe] 
ratios as well as higher SFRs at a fixed [Fe/H] abundance. Furthermore, the models with $\nu=3$ Gyr$^{-1}$
and $\lambda=9$ Gyr$^{-1}$ with the \citet{salpeter1955} and \citet{chabrier2003} IMFs predict always higher 
[$\alpha$/Fe] abundances than the models calculated with the IGIMF. 
This is due to the extremely low efficiency of formation of massive stars when the IGIMF is assumed 
for galaxies with very low SFRs.
\par The effect of changing the wind parameter 
$\lambda$ is much lower than varying the SF efficiency $\nu$. 
For a fixed value of the SF efficiency, the time of the onset of the galactic wind as well as 
the [Fe/H] ratio of the ISM at that epoch are always the same. 
So different values of the $\lambda$ parameter affect the chemical evolution only after the onset 
of the galactic wind. 
Once the wind has started, both the [$\alpha$/Fe] abundance ratios and the SFR decrease further.

\begin{table*}
\caption[sculptortable]{ \footnotesize{We report the differences produced in the averaged abundance ratios by adopting different sets of 
stellar yields and different IMFs. `Romano' stands for the yields adopted in \citet{romano2010}; `CL' stands for the yields of 
Chieffi and Limongi (private communication); `WW95' stands for the set of yields with \citet{woosley1995}, as described in the text. }}
\begin{tabular}[]{c | c c c c }
\hline
  & \small{$\Delta[\text{Si/Fe}]\pm\sigma$} & \small{$\Delta[\text{O/Fe}]\pm\sigma$} & \small{$\Delta[\text{Mg/Fe}]\pm\sigma$} & \small{$\Delta[\text{Ca/Fe}]\pm\sigma$} \\
\hline

\small{IMF: Salpeter} &  &  &  &  \\

\small{Romano/CL} & $0.033\pm0.007$ & $0.03\pm0.02$ & $0.22\pm0.03$ & $0.23\pm0.09$ \\

\small{WW95/CL} & $0.18\pm0.06$ & $0.12\pm0.04$ & $0.20\pm0.06$ & $0.33\pm0.13$ \\

\small{Romano/WW95} & $0.15\pm0.06$ & $0.13\pm0.02$ & $0.08\pm0.04$ & $0.10\pm0.05$ \\

\rule{0pt}{4ex}    

\small{IMF: Chabrier} & & & & \\

\small{Romano/CL} & $0.047\pm0.006$ & $0.05\pm0.03$ & $0.20\pm0.03$ & $0.25\pm0.10$ \\

\small{WW95/CL} & $0.18\pm0.06$ & $0.11\pm0.05$ & $0.18\pm0.06$ & $0.34\pm0.13$ \\

\small{Romano/WW95} & $0.13\pm0.06$ & $0.12\pm0.03$ & $0.07\pm0.04$ & $0.09\pm0.06$ \\

\rule{0pt}{4ex}

\small{IGIMF} & & & & \\

\small{Romano/CL} & $0.10\pm0.07$ & $0.28\pm0.24$ & $0.48\pm0.17$ & $0.11\pm0.07$ \\

\small{WW95/CL} & $0.13\pm0.05$ & $0.24\pm0.21$ & $0.48\pm0.30$ & $0.17\pm0.11$ \\

\small{Romano/WW95} & $0.14\pm0.08$ & $0.12\pm0.04$ & $0.12\pm0.07$ & $0.09\pm0.06$ \\

\hline

\end{tabular}
\label{table1}
\end{table*}

\par We have computed chemical evolution models of the Sgr dwarf with different stellar yield prescriptions, in order to 
provide a first estimate of the uncertainties due to the stellar yields. Our results are reported in Table \ref{table1}. 
We have tested different sets of stellar yields besides those of \citet{romano2010}, which we consider as the best in reproducing the solar vicinity 
abundance pattern. In particular, in addition to them, we have tested also:
\begin{enumerate}
 \item the yields of \citet[with the corrections suggested by \citealt{francois2004}]{woosley1995} 
 for massive stars, and the yields of \citet{vandenhoek1997} for low- and intermediate-mass stars;
 \item  the most recent yields from massive stars of the Chieffi and Limongi group (private communication), and 
 the yields of \citet{karakas2010} from low- and intermediate-mass stars.
\end{enumerate}
We find that the models with the \citet{romano2010} set of stellar yields agree very well in the predicted [O/Fe] and [Si/Fe] abundance ratios 
with the models which include the recent yields of the Chieffi and Limongi group. On the other hand, there is still quite a large uncertainty in the 
stellar yields of Mg and Ca, which affect the results of our models for these two chemical elements; in particular, the final results for [Mg/Fe] and [Ca/Fe] 
as a function of [Fe/H] may differ by almost $0.2$ dex.
\par In Table \ref{table1}, we explored also the combined effect of different IMF and stellar yield assumptions. 
We find that the effect of assuming different stellar yields is almost similar for [O/Fe] and [Si/Fe], if we assume the \citet{salpeter1955} or the \citet{chabrier2003} IMFs. 
On the other hand, when assuming the IGIMF, we find that our models become on average more influenced by the assumed stellar yields.

\section{Conclusions} \label{conclusions}

In this paper, we have tested the effects of different IMFs on the chemical evolution of Sgr
dwarf galaxy. In particular, we have considered the IGIMF of R14, which depends on the metallicity and SFR, 
and the invariant \citet{salpeter1955} and \citet{chabrier2003} IMFs. 
We have run several models by studying 
the effect of the various parameters, such as the efficiency of SF and the wind parameter. 
\par We have compared different scenarios for the production of Eu. In particular, we have considered 
the recent NSM scenario of \citet{matteucci2014} and the canonical scenario in which Eu is produced by core-collapse SNe. 
\par Finally, we have studied the effect of different stellar yield assumptions on the predicted abundance ratio patterns in this galaxy and we 
have explored also the combined effect of varying both the IMF and the stellar yield assumptions. 
\par In what follows, we summarize the main conclusions of our work.

\begin{enumerate}

\item The IGIMF tends to predict lower [$\alpha$/Fe] and [Eu/Fe] ratios in 
objects with low SFR than 
the classical \citet{salpeter1955} and \citet{chabrier2003} IMFs. In fact, 
in the case of the IGIMF, there is a deficiency in the formation of massive 
stars, which are the main contributors of the $\alpha$-elements.
The dependence of the IGIMF on the SFR is much stronger than 
that on the metallicity, which in fact could be neglected. Our results support the conclusion 
that the time-delay model is necessary to explain the trend of the [$\alpha$/Fe] and [Eu/Fe] ratios as a function of [Fe/H]; 
furthermore, the assumption of a truncated IMF such as the IGIMF provides a better qualitative agreement 
with the abundance ratio patterns observed in the Sgr galaxy, although 
both the data and the stellar yields that we assume in our models are still too uncertain to draw firm conclusions. 
It is worth recalling that the effect of changing the IMF consists mainly in shifting the  [$X$/Fe] (with $X$ the abundance of a generic element) 
curves up or down along the $Y$-axis, whereas the shape of the  [$X$/Fe] versus [Fe/H] curves is mainly determined by the lifetimes of the 
stellar producers of $X$ and Fe and by the star formation history (time-delay model).

\item The oxygen is the hydrostatic $\alpha$-element which is most sensitive to the cut-off in mass of the IMF, 
while the explosive $\alpha$-elements such as the silicon and the calcium are much less sensitive. 
So the hydrostatic over explosive $\alpha$-element abundance ratios can retain a well-defined signature of a truncated IMF, 
as suggested by \citet{mcwilliam2013}. The O and Si are among the chemical elements which are less affected by uncertainties 
relative to their stellar yields; the results of our models, in particular the comparison of the [O/Si] versus [Fe/H] relations predicted 
by our models with the \citet{mcwilliam2013} data, might support the idea that the IMF in the Sgr galaxy is truncated, with the IGIMF being the favourite 
among the different IMFs here explored. However, again, the data are still too uncertain to draw firm conclusions.

\item All our models with Eu coming from core-collapse SNe are not able to reproduce 
the [Eu/Fe] and [Eu/O] abundance ratios at the same time, unless the yields from stars in the range $8$-$10\,\text{M}_{\sun}$ are artificially increased 
by a factor of $\sim3$. When including the Eu produced by NSMs as the only source of this element, 
we are also able to well match the data, by assuming yields as suggested by recent calculations 
\citep{bauswein2014,wanajo2014,just2014}.  

\item Since NSMs, which are nowadays considered as more promising sites for Eu production, arise from stars which have an initial mass in a 
lower range than that of the most important oxygen producers, 
the hypothesis of \citet{mcwilliam2013} remains true. Furthermore, we confirm that in Sgr the truncation of the IMF might have played 
a relevant role in the [Eu/O] versus [Fe/H] relations, while the [Eu/Fe] versus [Fe/H] is due mainly to the time-delay model. 

\item By exploring the parameter space, and in particular by studying the effect of the star formation 
and galactic wind efficiencies, we found that the major role in determining the final abundance pattern 
in Sgr  galaxy is played by the star formation efficiency, while the wind parameter has only a small 
effect.

\item The IMFs considered here are all able to reproduce the present time observed total HI mass. 
On the other hand, the model with the IGIMF predicts final total stellar masses which are slightly 
larger than the ones predicted by the models with the classical IMFs. 
This is probably due to the delayed onset of the
galactic wind in IGIMF models, because of the reduced energetic feedback
from massive stars.  The galaxy forms stars for a longer period and thus a
large mass in long-living, low-mass stars can accumulate.

\item The present results can be useful to study also other dSphs, since the history of these galaxies 
is characterized by a low SFR, which implies a truncated IMF in the formalism of the IGIMF 
theory. In a forthcoming paper, we will discuss the chemical evolution of Fornax.

\end{enumerate}

Our last comment is that combining the chemical evolution models with the spectro-photometric ones might 
greatly help to better constrain the role and the effect of the IGIMF in the evolution of galaxies and 
that will be the subject of a forthcoming paper.

\section*{Acknowledgements}

FM acknowledges financial support from PRIN-MIUR~2010-2011 project 
`The Chemical and Dynamical Evolution of the Milky Way and Local Group Galaxies', prot.~2010LY5N2T. 
We thank D. Romano for useful suggestions and interesting discussions; we thank M. Chieffi and 
A. Limongi for kindly providing their stellar yields; we thank P. Kroupa for carefully reading the manuscript. 
Finally, we thank an anonymous referee for his/her suggestions which improved the clarity of this paper.

\bsp

\label{lastpage}

\end{document}